\documentclass[aps,prl,twocolumn,superscriptaddress]{revtex4-2}
\usepackage{graphicx} 
\usepackage[colorlinks=true, linkcolor=Blue, citecolor=Blue, urlcolor=Blue]{hyperref}
\usepackage{amssymb,amsmath}
\usepackage[usenames,dvipsnames]{xcolor}
\usepackage{txfonts}
\usepackage{bm}
\usepackage{soul}

\newcommand{\YZ}{\color{black}}
\newcommand{\YZZ}{\color{black}}
\setstcolor{blue}

\begin{document}
%===========
%
\title{Synchronizing Chaos with Imperfections}
\author{Yoshiki Sugitani}
\email[]{yoshiki.sugitani.0301@vc.ibaraki.ac.jp}
\affiliation{Department of Electrical and Electronic System Engineering, Ibaraki University, \\ 4-12-1 Nakanarusawa, Hitachi, Ibaraki 316-8511, Japan}
\author{Yuanzhao Zhang}
\email[]{yuanzhao@u.northwestern.edu}
\affiliation{Department of Physics and Astronomy, Northwestern University, Evanston, Illinois 60208, USA}
\affiliation{Center for Applied Mathematics, Cornell University, Ithaca, New York 14853, USA}
\author{Adilson E. Motter}
\email[]{motter@northwestern.edu}
\affiliation{Department of Physics and Astronomy, Northwestern University, Evanston, Illinois 60208, USA}
\affiliation{Northwestern Institute on Complex Systems, Northwestern University, Evanston, Illinois 60208, USA}
%
%\date{\today}
%
\begin{abstract}
Previous research on nonlinear oscillator networks has shown that chaos synchronization is attainable for identical oscillators but deteriorates in the presence of parameter mismatches. 
Here, we identify regimes for which the opposite occurs and show that  oscillator heterogeneity can 
synchronize chaos for conditions under which identical oscillators cannot. This effect is not limited to small mismatches and is observed for random oscillator heterogeneity on both homogeneous and heterogeneous network structures.
The results are demonstrated experimentally using networks of Chua's oscillators and are further supported by numerical simulations and theoretical analysis.
In particular, we propose a general mechanism based on heterogeneity-induced mode mixing that provides insights into the observed phenomenon.
Since individual differences are ubiquitous and often unavoidable in real systems, it follows that such imperfections can be an unexpected source of synchronization stability. 
\vspace{3mm}

\noindent DOI: \href{https://doi.org/10.1103/PhysRevLett.126.164101}{10.1103/PhysRevLett.126.164101} 
\end{abstract}

\maketitle

Synchronization in networks of chaotic oscillators is a remarkable phenomenon that is now well established theoretically and experimentally \cite{pecora2015synchronization}, 
%restrepo2006emergence,fujiwara2016synchronization
with implications for numerous biological and technological systems \cite{vanwiggeren1998communication,mosekilde2002chaotic,belykh2005synchronization,eroglu2017synchronisation}. Two conditions are generally assumed for this phenomenon to occur: (i) that the coupling strength be sufficiently large and (ii) that the oscillators be sufficiently identical. If the coupling is too weak, the oscillators evolve mostly independently from each other, and their trajectories tend to diverge due to sensitive dependence on initial conditions---a hallmark of chaos \cite{ott2002chaos}. On the other hand, if the oscillators are not sufficiently identical, their trajectories tend to diverge due to sensitive dependence on parameters---another hallmark of chaos \cite{ott2002chaos}---even if the initial conditions are exactly the same and the coupling is otherwise suitably strong. 

Previous analyses of synchronization of nonidentical chaotic oscillators have focused mainly on cluster synchronization \cite{belykh2003persistent,dahms2012cluster} and phase synchronization \cite{rosenblum1996phase,deshazer2001detecting,kiss2002phase,zhou2002noise,skardal2017optimal}.
For example, oscillator heterogeneity has been shown {\YZ to mediate relay synchronization \cite{fischer2006zero,vicente2008dynamical,gambuzza2016inhomogeneity} and to induce frequency locking by suppressing chaos \cite{braiman1995taming,lindner1997optimal,brandt2006synchronization,montaseri2016diversity}.}
Global chaos synchronization of nonidentical oscillators, 
on the other hand, has been explored mainly for strong coupling and small parameter mismatches \cite{rim2002routes,nishikawa2010network,sorrentino2011analysis,acharyya2012synchronization}, with an emphasis on the extent to which synchrony persists when the oscillators are slightly different \cite{Sun2009-ev,pereira2014towards}.
These previous results consistently show that global synchronization degrades as heterogeneity is increased.

A different body of work has shown that, for periodic oscillators, heterogeneity can in certain cases facilitate synchronization \cite{braiman1995disorder,bolhasani2015stabilizing,nishikawa2016symmetric,zhang2017nonlinearity,punetha2019clock,molnar2020network}.
%komin2011synchronization,
A natural question is then whether a similar effect would be possible for chaotic oscillators despite the fact that their dynamics exhibit sensitive dependence on parameters and that an invariant synchronization manifold no longer exists for nonidentical chaotic oscillators. This question is especially relevant in weak coupling regimes, in which synchronization is unstable for identical chaotic oscillators.
%The answer to this question is purportedly negative, at least for previously studied coupling regimes that lead to synchronization for their identical counterpart. But what about weak coupling regimes, in which synchronization is not possible for identical chaotic oscillators?

In this Letter, we experimentally demonstrate that oscillator heterogeneity can often enable synchronization of weakly coupled chaotic oscillators that would otherwise not synchronize. 
This result is established using Chua's oscillators diffusively coupled through their $x$ components, which leads to a semi-infinite stability region for identical oscillators. 
The robustness of the effect is confirmed by showing that it occurs consistently for {\it random} parameter heterogeneity and for {\it different} parameters (associated with temporal and state variable scales). 
The effect is also demonstrated across different network structures and is supported by simulations and theoretical analysis.
The role of {\it oscillator} heterogeneity is isolated by showing the persistence of the effect for structurally homogeneous networks of identically coupled oscillators. 
These results have immediate implications for real systems, where heterogeneity is ubiquitous. 
They also have foundational consequences for establishing an unanticipated 
relation between network coupling, oscillator heterogeneity, and sensitive dependence on initial conditions.
In particular, although condition (i) cannot be violated in isolation without causing desynchronization, our results show that the violation of (ii) (albeit  detrimental by itself) can mitigate the effect of infringing (i), and thus the synchronization of chaotic oscillators can persist if {\it both} (i) and (ii) are violated together.

We start by considering networks of $N$ diffusively coupled oscillators described by
\begin{equation}
\tau_i \dot{\bm{x}_i} = \bm{F}(\bm{x}_i) - k \sum_{j=1}^{N} L_{ij}\bm{H}(\bm{x}_{j}),\qquad i = 1,2,\dots,N,
\label{eq:non-d}
\end{equation}
where $\tau_i$ denotes the timescale and $\bm{x}_i$ is the state variable of the $i$th oscillator. 
The functions $\bm{F}$ and $\bm{H}$ describe the dynamics of a single oscillator and their interactions, respectively. 
The Laplacian matrix {\YZZ $\bm{L}=(L_{ij})=(\delta_{ij} \sum_\ell A_{i\ell} - A_{ij})$}, defined in terms of the adjacency matrix $\bm{A} = (A_{ij})$, represents the network structure. 
The parameter $k$ controls the coupling strength.

\begin{figure}[tb]
	\centering
	\includegraphics[width=8.6cm]{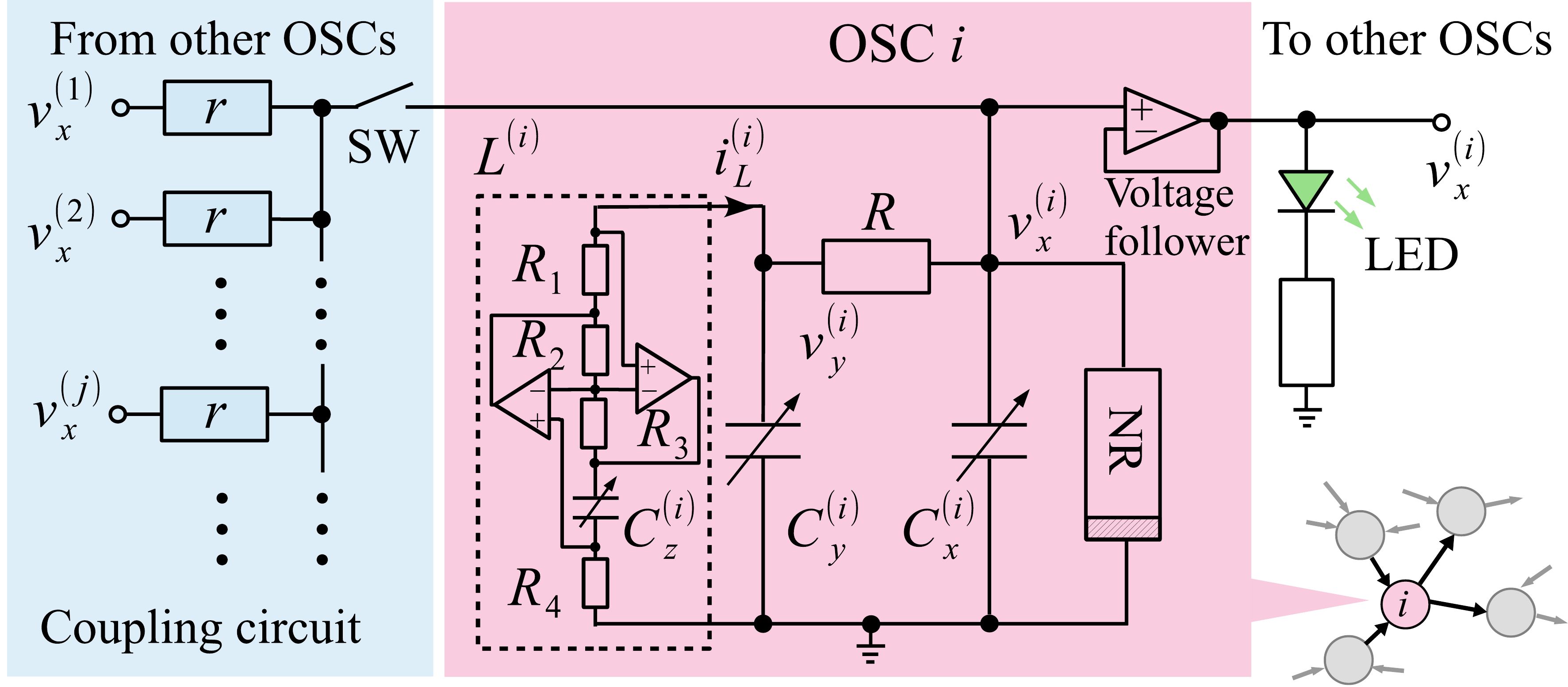}
	\caption{Circuit diagram of coupled Chua's oscillators in our experiment. The individual oscillators are coupled through their voltage $v_x^{(i)}$, and an LED is attached to each oscillator to visualize the voltage oscillations. The capacitors are tunable and
	control  the heterogeneity across the oscillators.
	\label{fig:circuit}}
\end{figure}

The oscillators and  coupling are implemented in our experiment using electrical circuits, as shown in Fig.~\ref{fig:circuit}.
The oscillators consist of $x$-coupled Chua's circuits \cite{Matsumoto1985-uq} modeled as
\begin{eqnarray}
\bm{F}(\bm{x})&=&
\begin{bmatrix}
\eta\{ y - x -g(x) \}\\
x-y+z\\
-y/\gamma
\end{bmatrix},
\;\;\bm{H}(\bm{x})=
\begin{bmatrix}
x\\
0\\
0
\end{bmatrix},\label{eq:F}\\
g(x) &=& bx + \frac{1}{2}(b-a)\left( |x-1| - |x+1| \right), 
\label{eq:g}
\end{eqnarray}
where $x$, $y$, and $z$ are the state variables and $\eta$, $\gamma$, $a$, and $b$ are parameters.
The variables correspond respectively to the voltages $v_x$ and $v_y$ across the capacitors $C_x$ and $C_y$ and the current $i_{L}$ through the inductor $L$ (which is implemented using a generalized impedance converter circuit). 
%{\YZ To mimic the voltage limit in the experiments, we also impose the boundary conditions $\dot{x}=0$ at $|x|=7$ and $\dot{y}=0$ at $|y|=3$ for Eqs.~(\ref{eq:non-d})-(\ref{eq:g}).}
The parameters $a$ and $b$ are determined by a nonlinear resistor (NR) with a piecewise linear characteristic made from op-amps (TL084) and resistors \cite{Kennedy1992-fl}. 
The tunable parameters of the oscillators are controlled through tunable capacitors.
The oscillators are coupled through the voltage $v_x$, 
where the directionality of the coupling is implemented using voltage followers. 
A light-emitting diode (LED) is attached to each oscillator so as to monitor the oscillation visually, with the diode turning on for $v_x>0$ and off for $v_x<0$.
Throughout the paper, we focus on the synchronization of $v_x$. 
The voltage $v_x^{(i)}$ for each oscillator is recorded by a computer using an analog-to-digital converter attached to the circuit.

The circuit parameters and variables are associated with the dimensionless quantities in Eqs.~(\ref{eq:non-d})--(\ref{eq:g}) as follows:
\begin{eqnarray}
\tau_i&=&\frac{C_y^{(i)}}{\bar{C}_y},\;\;k=\eta\frac{R}{r},\;\; \eta=\frac{C_y^{(i)}}{C_x^{(i)}},\;\; \gamma= 
\frac{C_z^{(i)}R_1 R_3 R_4}{C_y^{(i)}R^2 R_2},\notag\\
a&=&m_1R,\;\;b=m_0R,\;\;   x^{(i)}=\frac{v_x^{(i)}}{B_p},\;\;y^{(i)}=\frac{v_y^{(i)}}{B_p}, \;\;z^{(i)}=\frac{i_{L}^{(i)} R}{B_p},\notag
\end{eqnarray}
where $\bar{C}_y=\frac{1}{N}\sum_{i=1}^{N}C_y^{(i)}$, $m_1$ and $m_0$ are determined by the %resistors in 
NR, and $B_p$ depends on both the saturation voltage of the op-amps and the resistors connected to them \cite{Kennedy1992-fl}. 
The dimensionless time used in Eq.~(\ref{eq:non-d}) is defined as $t' = t/(R\bar{C}_y)$ and,
without loss of generality, it follows that the mean timescale is $\bar{\tau}=\frac{1}{N}\sum_{i=1}^{N} \tau_i\equiv 1$ (this condition is also imposed in our simulations and analysis).  
Unless noted otherwise, the oscillator parameters are fixed at $\eta = 10$,  $\gamma = 0.056$,  $a = -1.44$, and $b=-0.72$,
which gives rise to a double-scroll chaotic attractor in the absence of coupling \cite{masamura2020experimental}.
% [Masamura, S., Iwamoto, T., Sugitani, Y., Konishi, K. & Hara, N. Experimental investigation of amplitude death in delay-coupled double-scroll circuits with randomly time-varying network topology. Nonlinear Dyn. 99, 3155Ð3168 (2020)]
These parameters are realized in the experiment by setting 
$R_1 = R_2 = R_3 = 1 \,\text{k}\Omega$, $R=R_4=1.8 \,\text{k}\Omega$, $\bar{C}_y = 5.7 \,{\mu}$F,
$B_p=1.3 \,\text{V}$, $m_0=-0.4 \,\text{m}\Omega^{-1}$, and $m_1=-0.8 \,\text{m}\Omega^{-1}$,
and by keeping the capacitance ratios as
\begin{equation}
C_y^{(i)} = 10 C_x^{(i)}, \; C_z^{(i)} = C_x^{(i)},
\label{eq:relation_C}
\end{equation}
which ensure the same $\eta$ and $\gamma$ values across all oscillators.
The capacitors $C_y^{(i)}$, which control the timescales $\tau_i$, are tuned to vary the heterogeneity among the oscillators, 
and the resistors $r$ are changed to modify the coupling strength $k$ (see Supplemental Material \cite{sm} for details).

\begin{figure}[t]
	\centering
	\includegraphics[width=8.6cm]{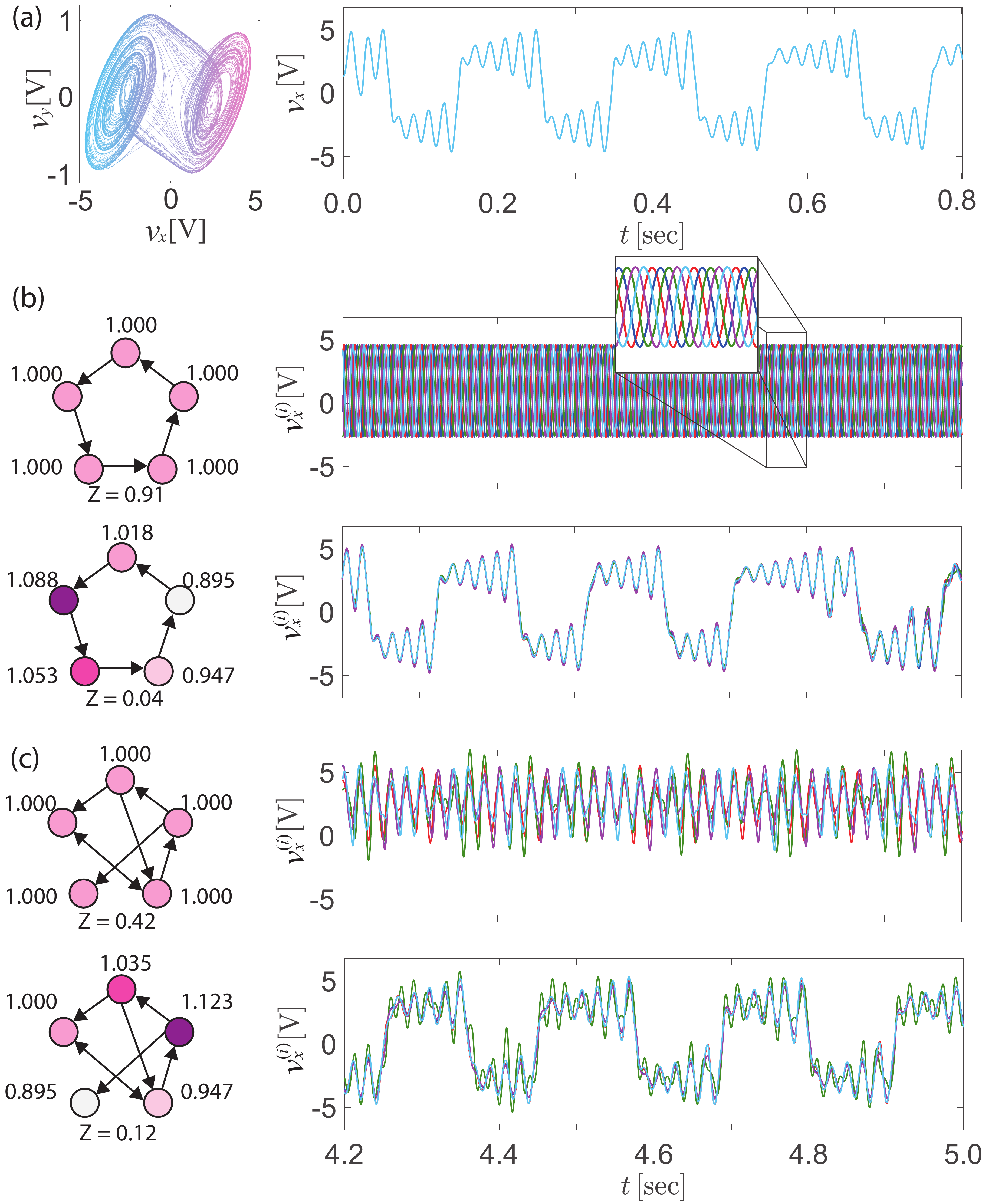}
	\label{fig:ring}
	\vspace{-5mm}
	\caption{Chaos synchronization induced by random oscillator heterogeneity.
	(a) Double-scroll chaotic attractor constructed from the experimental time series of the voltages $v_x$ and $v_y$ of a single uncoupled oscillator, and the time series for $v_x$ shown separately.
	(b, c) Corresponding experimental time series for oscillators in a directed ring for $k=8.18$ (b) and in a random network for $k=5$ (c). 
	Left: network structures and synchronization errors $Z$, where each node is labeled with its timescale $\tau_i$. 
	Right: time series after the initial transient (colored by oscillator) for initial conditions close to the synchronous state, showing that chaos synchronization is stable in the heterogeneous system but not in the homogeneous one.
	In particular, the heterogeneous systems both achieve low synchronization error and preserve qualitative properties of the original chaotic attractor.
	Videos of the time series in (b), tracked by LEDs attached to the oscillators, are included in the Supplemental Material~\cite{sm}.
	\label{fig:time}}
\end{figure}

We first analyze in Fig.~\ref{fig:time}(b) the experimental time series of $v_x^{(i)}$ for a directed ring network of five oscillators, where the coupling strength is below the synchronization transition threshold for the identical oscillators.
The upper panel confirms that, for homogeneous timescales, the trajectories of the oscillators diverge from each other and the system moves away from the synchronous chaotic state.
In the lower panel, we introduce a random perturbation to the timescales, as indicated on the network image. 
Although the synchronization manifold $\bm{x}_1=\dots=\bm{x}_N$ is no longer invariant, the heterogeneous system remains closely synchronized for the duration of the experiment and, collectively, exhibits double-scroll chaotic dynamics comparable to those of the uncoupled oscillators [Fig.~\ref{fig:time}(a)].
Figure~\ref{fig:time}(c) shows a similar result for a random network with nonuniform indegrees.
Once again, for the subcritical coupling strength considered, synchronization is lost in the homogeneous system but persists in the heterogeneous system.

The degree of synchronization in our experiment is further quantified by calculating the synchronization error $Z$ based on the measured voltages $v_x^{(i)}$. 
We first define $e_{v_x}=\left\langle \sqrt{\frac{1}{N}\sum_{i=1}^{N} \left[v_x^{(i)}(t)-\bar{v}_x(t)\right]^2 } \, \right\rangle$,
%\begin{equation}
%e_{v_x}=\left\langle \sqrt{\frac{1}{N}\sum_{i=1}^{N} \left[v_x^{(i)}(t)-\bar{v}_x(t)\right]^2 } \, \right\rangle,
%\label{eq:Z}
%\end{equation}
where $\bar{v}_x(t)$ is the average of $v_x^{(i)}(t)$ over the oscillators at time $t$, and $\langle\cdot\rangle$ denotes the time average over a period of $5\,$seconds after the initial transient. 
To facilitate comparison between different conditions, we normalize $e_{v_x}$ by the standard deviation of $v_x$ calculated over the $5$-second trajectory segment of all oscillators.
%, which includes $5$ seconds of trajectory for each $i$.
The synchronization error $Z$ is then calculated as the average of the normalized $e_{v_x}$ over the experimental trials (taken to be $5$ in Fig.~\ref{fig:time}).
In Fig.~\ref{fig:time}(b), for example, the synchronization error for the homogeneous system is $Z=0.91$, whereas for the heterogeneous system it is $Z=0.04$.

\begin{figure}[tb]
\centering
\includegraphics[clip,width=\columnwidth]{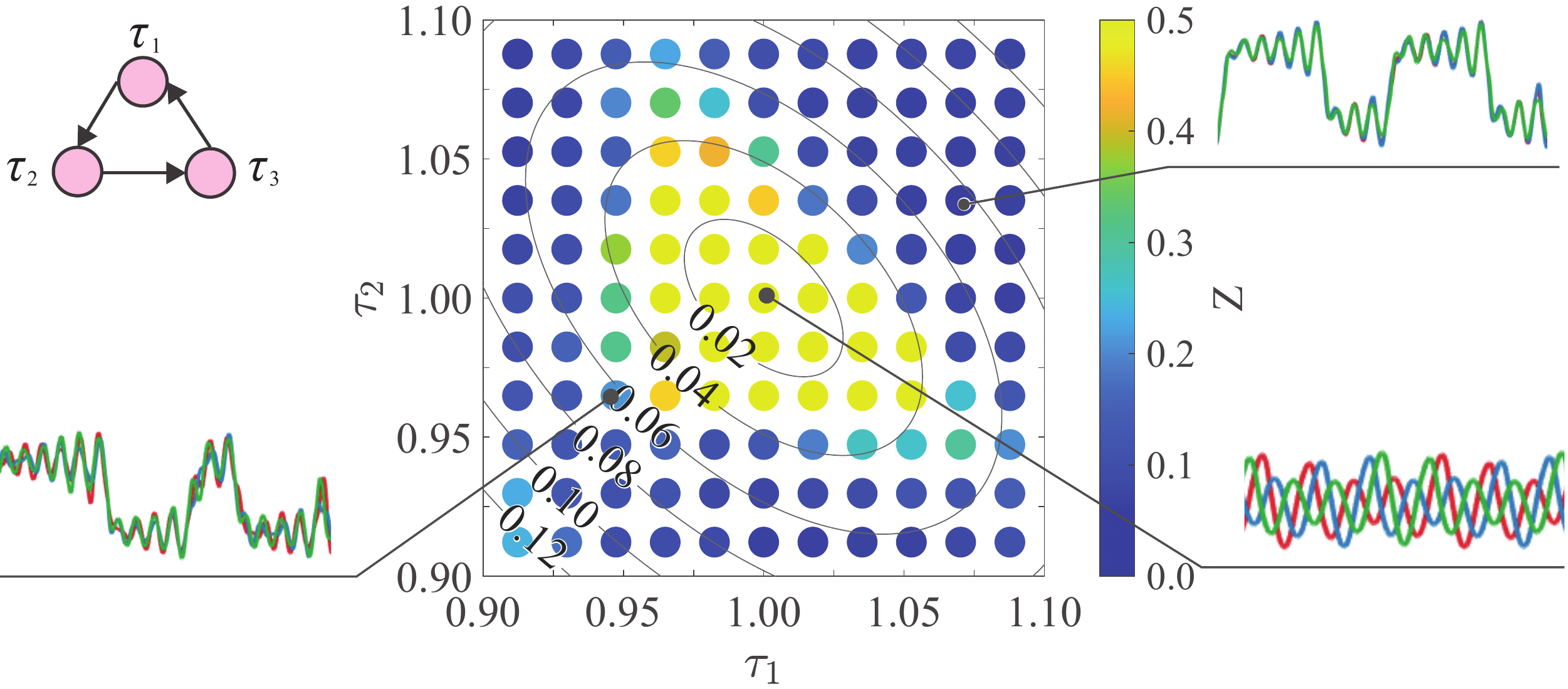}
\vspace{-5mm}
\caption{Experimental results for a directed ring network of three oscillators. 
The diagram shows synchronization error $Z$ measured from the experiments, where the contour lines indicate the standard deviation $\sigma$ among $\tau_i$. 
The outsets show representative examples of the time series of $v^{(i)}_x$ for the corresponding parameters.
Each data point is determined from $10$ experimental trials, with each trial starting from a random initial condition.
The coupling strength is set to $k=5$.
\label{fig:N3_d1_d2}}
\end{figure}

To explore the effect of random heterogeneity more systematically, we focus on a minimal system of three circuit oscillators.
Figure~\ref{fig:N3_d1_d2}
%(a) 
shows the synchronization error $Z$ in the $(\tau_1,\tau_2)$ parameter space, for $\tau_3$ satisfying $\tau_3=3-\tau_1-\tau_2$.
The center of the image $(\tau_1,\tau_2,\tau_3)=(1,1,1)$ corresponds to the homogeneous system, which is characterized by a large synchronization error.
As we move away from the center, $Z$ eventually decreases to a value close to zero in all directions. 
The contours in the figure show the standard deviation $\sigma$ among the $\tau_i$. 
The synchronization error $Z$ decreases abruptly when $\sigma$ becomes larger than approximately $0.05$, which indicates that oscillator heterogeneity consistently promotes chaos synchronization in our system under the given conditions.
Our numerical simulations for the same network and parameters using the model in Eqs.~(\ref{eq:non-d})--(\ref{eq:g}) show a qualitatively similar transition to synchronization as the heterogeneity is increased, further supporting the experimental findings.
%These simulation results and additional experimental results are presented in the Supplemental Material \cite{sm}, 
These results are presented in the Supplemental Material \cite{sm}, where we also show that heterogeneity in the parameter $\gamma_i$ stabilizes synchronization equally well, and that the phenomenon is also observed experimentally for Chua's circuits exhibiting single-scroll chaotic dynamics.

Having shown experimentally that there are scenarios under which random parameter heterogeneity facilitates synchronization, we now present a theory identifying the general mechanism behind this phenomenon.
For a network of identical oscillators, the variational equation governing the synchronization stability has the following form:
\begin{equation}
    \delta\dot{\bm{X}} = \left[ \bm{I}_N \otimes {\rm D}\bm{F}(\bm{x}) - k \bm{L} \otimes {\rm D}\bm{H}(\bm{x}) \right] \delta\bm{X},
  \label{eq:var}
\end{equation}
where {\YZZ $\delta\bm{X} = (\bm{x}_1^\intercal-\bm{x}^\intercal,\dots,\bm{x}_N^\intercal-\bm{x}^\intercal)^\intercal$} is the perturbation vector, $\bm{I}_N$ is the $N\times N$ identity matrix, $\bm{x}$ is the synchronization trajectory, and $\otimes$ is the Kronecker product. 
For simplicity, we assume the Laplacian matrix $\bm{L}$ is diagonalizable and, consequently, Eq.~(\ref{eq:var}) can be decoupled by applying a coordinate transformation $\bm{Q}$ for which $\bm{Q}^{-1}\bm{L}\bm{Q}=\bm{\lambda} = \text{diag}(\lambda_1,\dots,\lambda_N)$ \cite{pecora1998master}.
This gives rise to $N$ lower-dimensional equations of the same form, each corresponding to an independent perturbation mode $\bm{\xi}_i$:
\begin{equation}
	\dot{\bm{\xi}}_i = \left[ \text{D} \bm{F}(\bm{x}) - k\lambda_i {\rm D}\bm{H}(\bm{x}) \right] \bm{\xi}_i, \qquad i = 1,2,\dots,N.
	\label{eq:msf}
\end{equation}
When oscillators become nonidentical, ${\rm D}\bm{F}$ (and ${\rm D}\bm{H}$ in the case of heterogeneous timescales $\tau_i$) becomes different for each oscillator.

For small heterogeneity, assuming the change in the synchronization trajectory is negligible, the mismatches in ${\rm D}\bm{F}$ and ${\rm D}\bm{H}$ introduce a perturbation matrix $\bm{\Delta}(\bm{x})$ in Eq.~(\ref{eq:var}):
\begin{equation}
    \delta\dot{\bm{X}} = \left[ \bm{I}_N \otimes {\rm D}\bm{F}(\bm{x}) - k\bm{L} \otimes {\rm D}\bm{H}(\bm{x}) + \bm{\Delta}(\bm{x}) \right] \delta\bm{X}.
  \label{eq:var2}
\end{equation}
Now, when we apply the transformation matrix $\bm{Q}$ to Eq.~(\ref{eq:var2}), we get 
\begin{equation}
    \dot{\bm{\xi}} = \left[ \bm{I}_N \otimes {\rm D}\bm{F}(\bm{x}) -  k\bm{\lambda} \otimes {\rm D}\bm{H}(\bm{x}) + \widetilde{\bm{\Delta}}(\bm{x}) \right] \bm{\xi},
  \label{eq:var3}
\end{equation}
where $\bm{\xi} = (\bm{\xi}_1^\intercal,\bm{\xi}_2^\intercal,\dots,\bm{\xi}_N^\intercal)^\intercal$, and $\widetilde{\bm{\Delta}}$ is $\bm{\Delta}$ under the new coordinates. 
Dividing $\tilde{\bm{\Delta}}$ into $N\times N$ blocks of equal size,
%$\tilde{\bm{\Delta}} = (\tilde{\bm{\Delta}}_{ij})$,
\begin{equation}
	\tilde{\bm{\Delta}} = 
    \begin{pmatrix}
    \tilde{\bm{\Delta}}_{11} & \tilde{\bm{\Delta}}_{12} & \cdots & \tilde{\bm{\Delta}}_{1N} \\
    \tilde{\bm{\Delta}}_{21} & \tilde{\bm{\Delta}}_{22} & \cdots & \tilde{\bm{\Delta}}_{2N} \\
    \vdots  & \vdots  & \ddots & \vdots  \\
    \tilde{\bm{\Delta}}_{N1} & \tilde{\bm{\Delta}}_{N2} & \cdots & \tilde{\bm{\Delta}}_{NN} 
    \end{pmatrix},
\end{equation}
Eq.~(\ref{eq:var3}) can be written as
\begin{equation}
	\dot{\bm{\xi}}_i = \left[ \text{D} \bm{F}(\bm{x}) - k\lambda_i {\rm D}\bm{H}(\bm{x}) \right] \bm{\xi}_i + \sum_{j} \tilde{\bm{\Delta}}_{ij}(\bm{x}) \bm{\xi}_j. %\qquad i = 1,2,\dots,N.
	\label{eq:msf_mixing}
\end{equation}
Crucially, $\tilde{\bm{\Delta}}$ is generally non-block-diagonal in these coordinates.
Comparing Eq.~(\ref{eq:msf_mixing}) with Eq.~(\ref{eq:msf}), it is clear that one of the main effects of parameter mismatches is the introduction of interdependence among the (originally independent) perturbation modes due to off-diagonal blocks in matrix $\tilde{\bm{\Delta}}$.
%and it is clear that there is now interdependence among the perturbation modes due to the presence of parameter mismatches.
%Here, each perturbation mode $\bm{\xi}_i$ corresponds to a $d$-dimensional subspace that is invariant under the evolution of the variational equation, and $d$ is the dimension of a single oscillator.

%Consequently, we need to go beyond Eq.~\eqref{eq:var_eq} to explain our experimental and numerical findings.
%One key assumption of the extended MSF is that the perturbation modes still evolve independently from each other.
Consequently, in the presence of random oscillator heterogeneity, 
%however small, 
the perturbation modes $\bm{\xi}_i$ are no longer mutually independent and start to ``mix'' with each other.
%In order to reduce the complexity of the analysis, 
This effect is generally not captured by previous efforts to analyze synchronization in networks of heterogeneous oscillators.
For example, the extended master stability function (MSF) assumes that, under small parameter mismatch, the variational equation can still be decoupled into $N$ independent equations, each corresponds to one of the original perturbation modes $\bm{\xi}_i$ \cite{Sun2009-ev}.
A simple analysis shows that, within the framework of the extended MSF, the Lyapunov exponents of each perturbation mode do not depend on oscillator heterogeneity (see Supplemental Material \cite{sm} for details).
This motivates us to go beyond the extended MSF in our theory, and it also justifies our choice to focus on the effect of heterogeneity-induced mode mixing.
%the $N$ original perturbation modes 
%$\bm{\xi}_i$ still evolve independently from each other.
%Our decision to focus on mixing is also motivated by the fact that, within the framework of extended MSF, the Lyapunov exponents associated with each perturbation mode do not depend on oscillator heterogeneity (see Supplemental Material \cite{sm} for details).
%More precisely, due to parameter mismatches, the full variational equation can no longer be decoupled into $N$ equations by diagonalizing the Laplacian matrix.
%, the invariance of Lambda becomes very useful (see below).

To intuitively understand how the interdependence among perturbation modes affects stability, consider a network of identical oscillators that is weakly unstable, which typically has one unstable and $N-2$ stable transverse perturbation modes.
%After introducing oscillator heterogeneity, from the extended MSF we know that (up to the first order) the Lyapunov exponents associated with each perturbation mode do not change.
After introducing oscillator heterogeneity, the invariant subspaces of the variational equation are destroyed and the perturbation modes become interdependent. 
A perturbation vector initially aligned with the weakly unstable direction will spend time in other (more stable) directions, which can potentially lead to an averaging effect that stabilizes against all possible perturbations.
%For more details on how parameter mismatches generate interdependence among perturbation modes, see Supplemental Material \cite{sm}.

\begin{figure}[tb]
\centering
\includegraphics[width=0.99\columnwidth]{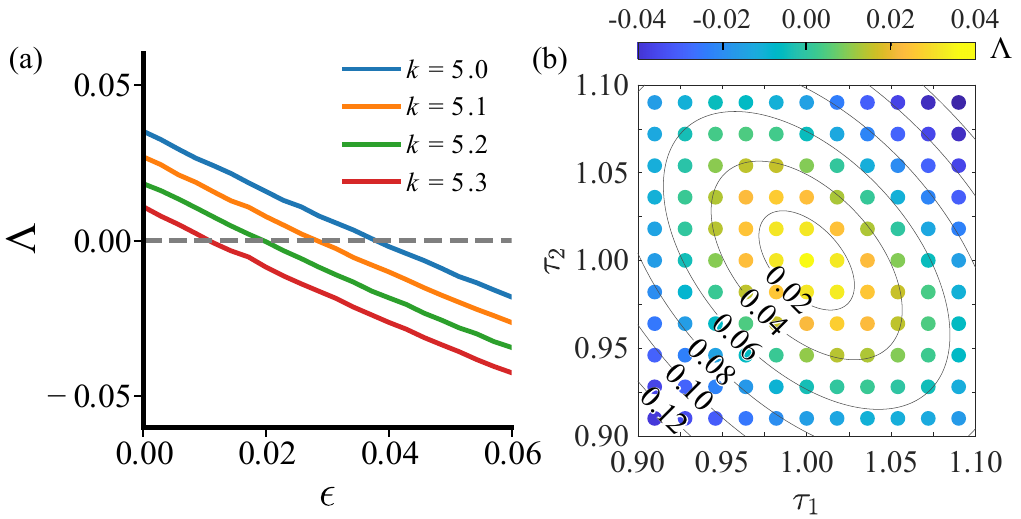}
\vspace{-4mm}
\caption{Theoretical predictions based on heterogeneity-induced mode mixing for the system considered in Fig.~\ref{fig:N3_d1_d2}.
(a) Dependence of the largest Lyapunov exponent $\Lambda$ on the mixing coefficient $\epsilon$ for a range of values of coupling strength $k$ in Eq.~\eqref{eq:mixing}. 
%In all cases, the stability is predicted to improve as $\epsilon$ is increased from zero.
(b) Dependence of $\Lambda$ on $\tau_i$ predicted using Eq.~(\ref{eq:msf_mixing}) under the approximation of an invariant synchronization manifold. 
The theoretical prediction qualitatively agrees with the experimental results in Fig.~\ref{fig:N3_d1_d2}.
}
\label{fig:MLE}
\end{figure}

% YZ: further justify not including $xi_1$
To heuristically model this effect, we introduce a mixing coefficient $\epsilon$ that controls the mixing rate between different transverse perturbation modes $\bm{\xi}_i$.
To keep the formalism simple, we assume that $\epsilon$ is a constant and that heterogeneity causes each transverse mode to ``leak'' into all other transverse directions at the same rate $\epsilon$.
The model can then be described as
%\begin{equation}
%	\dot{\bm{\xi}}_i = \left[ \text{D}_{\bm{x}} F(\bar{\bm{x}},\bar{p}) - \alpha_i {\rm D}_{\bm{x}}H(\bar{\bm{x}},\bar{p}) \right] \bm{\xi}_i - \epsilon (N-1)\bm{\xi}_i + \epsilon\sum_{j=2}^N \bm{\xi}_j.
%	\label{eq:mixing}
%\end{equation}
%\begin{equation}
%	\dot{\bm{\xi}}_i = \left[ \text{D} \bm{F}(\bm{x}) - k\lambda_i {\rm D} \bm{H}(\bm{x}) \right] \bm{\xi}_i - \epsilon (N-1)\bm{\xi}_i + \epsilon\sum_{j=2}^N \bm{\xi}_j.
%	\label{eq:mixing}
%\end{equation}
\begin{equation}
	\dot{\bm{\xi}}_i = \left[ \text{D} \bm{F}(\bm{x}) - k\lambda_i {\rm D} \bm{H}(\bm{x}) \right] \bm{\xi}_i + \epsilon\sum_{j=2}^N (\bm{\xi}_j-\bm{\xi}_i), \quad i=2,\dots,N,
	\label{eq:mixing}
\end{equation}%
where {\YZZ we used that $\epsilon\sum_{j=2}^N (\bm{\xi}_j-\bm{\xi}_i) = -\epsilon(N-1)\bm{\xi}_i + \epsilon\sum_{j=2}^N \bm{\xi}_j$.
Here, $-\epsilon(N-1)\bm{\xi}_i$ represents mode $i$ leaking out of direction $i$ and $\epsilon\sum_{j=2}^N \bm{\xi}_j$ represents the other transverse modes leaking into direction $i$.}
In general, one can expect that the mixing coefficient $\epsilon$ will grow with the magnitude of parameter mismatches in the system. 
%The excluded index $i=1$ corresponds to the mode parallel to the synchronization manifold of the homogeneous system.
To eliminate confounding factors, in Eq.~\eqref{eq:mixing} we made the approximation that the synchronization manifold remains invariant, and we use the same synchronization trajectory as in Eq.~\eqref{eq:msf} for calculating the largest (transverse) Lyapunov exponent $\Lambda$.
Thus, the index $i$ in Eq.~\eqref{eq:mixing} runs from $2$ to $N$ in order to include only modes transverse to the synchronization manifold.
%and does not include the parallel perturbation $\bm{\xi}_1$, which is parallel to the synchronization manifold.
%Finally, we omit the $\beta$ term from Eq.~\eqref{eq:var_eq} since it does not affect stability.
%In general, the coefficients in front of $\bm{\xi}_j$ are $d\times d$ matrices (where $d$ is the dimension of a single oscillator), whose norms grow with the magnitude of parameter mismatches in the system.
%The exact correspondence between the mixing coefficient and the oscillator heterogeneity depends on the system details and is left for future research.

Figure~\ref{fig:MLE}(a) shows $\Lambda$ as a function of $\epsilon$ for different values of $k$ in Eq.~\eqref{eq:mixing} and for $\lambda_i$ corresponding to the directed ring network used in Fig.~\ref{fig:N3_d1_d2}.
%This system is especially interesting because all the transverse perturbation modes have the same Lyapunov exponents and become unstable at the same time.
As we increase the mixing coefficient $\epsilon$, 
%(by increasing heterogeneity), 
$\Lambda$ decreases and eventually becomes negative for all relevant $k$.
In Supplemental Material \cite{sm}, we further support our theory by providing an explicit correspondence between the parameter mismatches in $\tau_i$ and the mixing matrix $\tilde{\bm{\Delta}}$. 
This enables us to explicitly compare the theoretical prediction [Fig.~\ref{fig:MLE}(b)] with the experimental results, confirming a good agreement between the two.
Thus, mode mixing is the dominant contributor to the {\YZ improved} synchronization observed in our experiments and constitutes a general mechanism through which parameter mismatches in coupled oscillators can facilitate synchronization.
%The effect of other factors caused by parameter mismatches, such as the shift in the synchronization trajectory, are left for future investigation.

It is natural to ask: for what classes of systems and under what conditions can we expect to observe heterogeneity-induced synchronization? 
The answer to this question lies on the balance between two competing effects of heterogeneity on synchronization. 
On the one hand, we have shown that heterogeneity {\YZ tends} to improve stability by introducing mixing among the perturbation modes. 
On the other hand, the synchronization manifold often becomes ``fuzzy'' (i.e., the trajectory deviates from identical synchronization) when oscillators are nonidentical, and the permissible synchronization states deteriorate as heterogeneity is increased. % without considering stability
For example, Ref.~\cite{Sun2009-ev} shows that when synchronization is stable, 
%the deviation from perfect synchrony 
the synchronization error increases linearly with oscillator heterogeneity. 
Thus, whether coherence in a network increases or decreases with heterogeneity depends on which of the two competing effects dominates, rather than on the details of the node dynamics and network structure.
%For example, heterogeneity is not expected to improve synchronization in linear systems (i.e., when $\mathrm{D}\bm{F}$ and $\mathrm{D}\bm{H}$ in the variational equation are constant matrices).
%In such systems, the variational equation can be fully decoupled even when oscillators are nonidentical (i.e., independent perturbation directions continue to exist in heterogeneous systems), and thus the mixing mechanism 
%does not directly apply.
%is not applicable.
%This could explain why similar phenomena have not been observed for Kuramoto oscillators.}
%{\YZZ On the other hand, for nonlinear systems with time-dependent $\mathrm{D}\bm{F}$ or $\mathrm{D}\bm{H}$, mode mixing (and thus stability improvement) can generally be expected in the presence of oscillator heterogeneity.}

%Based on the discussion above, our finding does not depend on the details of the node dynamics and network structure, but rather on a competition between mode mixing and loss of invariance of the synchronization manifold as heterogeneity is increased.
As a result, the phenomenon of heterogeneity-induced synchronization is expected to apply to a broad class of networked dynamical systems.
To further support this expectation, in the Supplemental Material \cite{sm} we show that oscillator heterogeneity can also lead to dramatic improvement of synchronization in non-autonomous systems describing driven pendulum arrays.
As in the networks of Chua's oscillators, in this case too the synchronization trajectory remains qualitatively similar to the otherwise unstable synchronization trajectory of the homogeneous system, despite the absence of an invariant synchronization manifold.
Future theoretical work will have the opportunity to further characterize the tradeoff between synchronization stability and the ``deformation'' of the synchronization manifold in networks of heterogeneous oscillators, especially for large heterogeneity.
Our demonstration that parameter heterogeneity can enable rather than halt synchronization has several implications. 
In particular, it completes a full circle in revealing the interplay between chaos and coupling interactions. 
%\cite{pecora1990synchronization}. 
%wang2002synchronization,belykh2004blinking,flunkert2010synchronizing
Early work on synchronization between coupled oscillators showed that sufficiently strong coupling can mitigate sensitive dependence on initial conditions.
By demonstrating that oscillator heterogeneity can enable synchronization below the synchronization transition threshold of identical oscillators, this work shows that, 
%even weak coupling can mitigate sensitive dependence on parameter assignment---and thus on initial conditions---and lead to convergence rather than divergence between the trajectories of mismatched systems. 
%Thus, our findings show that 
despite the sensitive dependence on oscillator parameters, 
parameter heterogeneity can reduce the effective coupling threshold for synchronization.
In man-made systems, this {\YZ suggests} that experimental imperfections may become an unexpected source of synchronization stability. 
In natural systems that rely on synchronization, it also suggests the possibility of observed mismatches being a result of evolutionary pressure that favors synchronization.

\begin{acknowledgments}
{\YZ This work was supported by JSPS KAKENHI Grant No.\ JP17K12748 (Y.S.), ARO Grant No.\ W911NF-19-1-0383 (A.E.M. and Y.Z.), and a Schmidt Science Fellowship (Y.Z.).}

Y.S. and Y.Z. contributed equally to this work.
\end{acknowledgments}

\bibliography{net_dyn}
\end{document}